\documentclass[fleqn,12pt,twoside]{article}
\usepackage{espcrc1}

\usepackage{epsf}

\newcommand{\AmS}{{\protect\the\textfont2
  A\kern-.1667em\lower.5ex\hbox{M}\kern-.125emS}}

\def\simgt{\,\rlap{\lower 3.5 pt\hbox{$\mathchar \sim$}}\raise 1pt \hbox {$>$}\,}
\def\simlt{\,\rlap{\lower 3.5 pt\hbox{$\mathchar \sim$}}\raise 1pt \hbox {$<$}\,}

\title{
\vspace*{-60pt}
{\normalsize \hfill {\sf UTHEP-448}} \\
\vspace*{-2pt}
{\normalsize \hfill {\sf October, 2001}} \\
\vspace*{40pt}
       Thermodynamic properties of QCD with two flavors of 
       Wilson-type lattice quarks
\thanks{Talk presented at {\it Statistical QCD}, Aug. 26--30, 2001,
Bielefeld, Germany.}
      }
\author{Kazuyuki Kanaya%
	\address[MCSD]{Institute of Physics, 
	University of Tsukuba, 
        Tsukuba, Ibaraki 305-8571, Japan} 
        for the CP-PACS Collaboration%
	}
       
\begin{document}

\maketitle

\begin{abstract}
I report on a study of finite temperature QCD 
by the CP-PACS Collaboration 
toward a precise determination of the equation of state 
with dynamical u,d quarks. 
Based on a systematic simulation using improved Wilson-type quarks
on lattices with temporal size $N_t=4$ and 6,
the energy density and pressure are calculated
as functions of temperature and renormalized light quark mass 
in the range $T/T_c \approx 0.7$--2.5 and 
$m_{\rm PS}/m_{\rm V} = 0.65$--0.95. 
Results for $N_t=4$ are found to contain significant scaling violations, 
while results for $N_t=6$ are suggested to be 
not far from the continuum limit. 
On the other hand, the quark mass dependence in the EOS 
turned out to be small for $m_{\rm PS}/m_{\rm V} \simlt 0.8$. 
\end{abstract}

\section{INTRODUCTION}

Theoretical understanding on the nature of the finite temperature QCD
phase transition and its thermodynamic properties is indispensable
in extracting an unambiguous signal of quark-gluon plasma production 
from heavy-ion collision experiments. 
In particular, the equation of state (EOS) of quark-gluon plasma 
belongs to the most basic information needed. 
Here, 
numerical simulations of lattice QCD provides us with the only systematic 
way to calculate these quantities directly from the first principles of QCD 
\cite{Kanaya98}. 

In this paper, I present our study of finite temperature QCD 
on the lattice with dynamical u and d quarks, 
focusing on the efforts toward a precise determination of the EOS 
\cite{ourPhase,ourEOS}.
Calculations are made on the CP-PACS computer, a dedicated parallel 
computer developed at the University of Tsukuba in 1996 
by physicists and computer scientists \cite{cp-pacs}.
With 2048 node processors the peak performance of the CP-PACS is 614.4 GFLOPS.
Intensive lattice QCD calculations on the CP-PACS have 
clarified the existence of quenched errors in the traditional calculations 
of light hadron spectrum in the quenched approximation \cite{CPPACSquench}, 
and opened the way towards precise full QCD calculations of hadronic 
quantities through the first systematic study of the light hadron 
spectrum with dynamical u,d quarks \cite{CPPACSfull}. 
Our simulations of finite temperature QCD are based on 
the experiences of these zero-temperature studies.

\section{FINITE TEMPERATURE QCD ON THE LATTICE}
\label{sec:lattice}

\begin{figure}[tb]
\centerline{
\epsfxsize=8cm \epsfbox{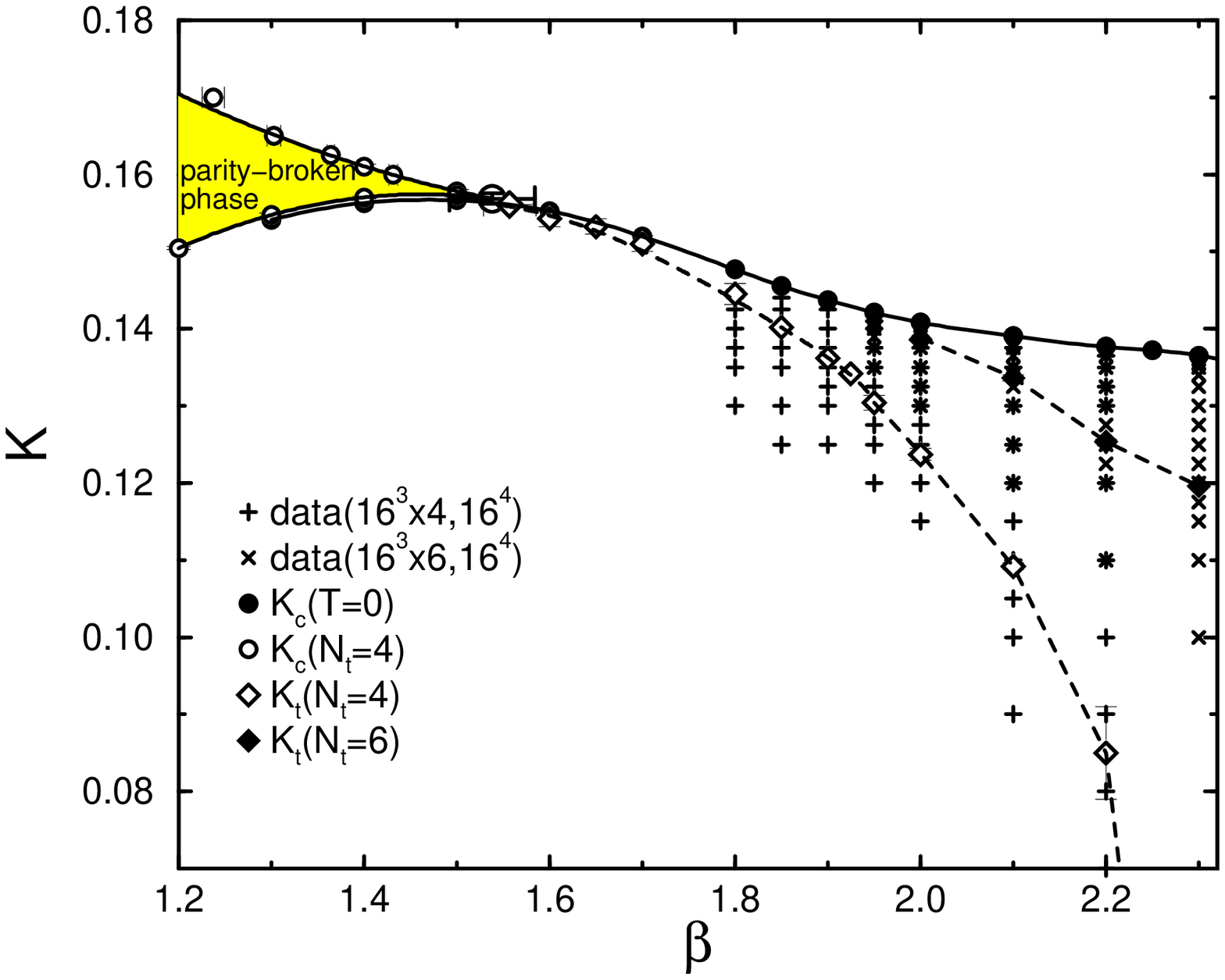}
\epsfxsize=8cm \epsfbox{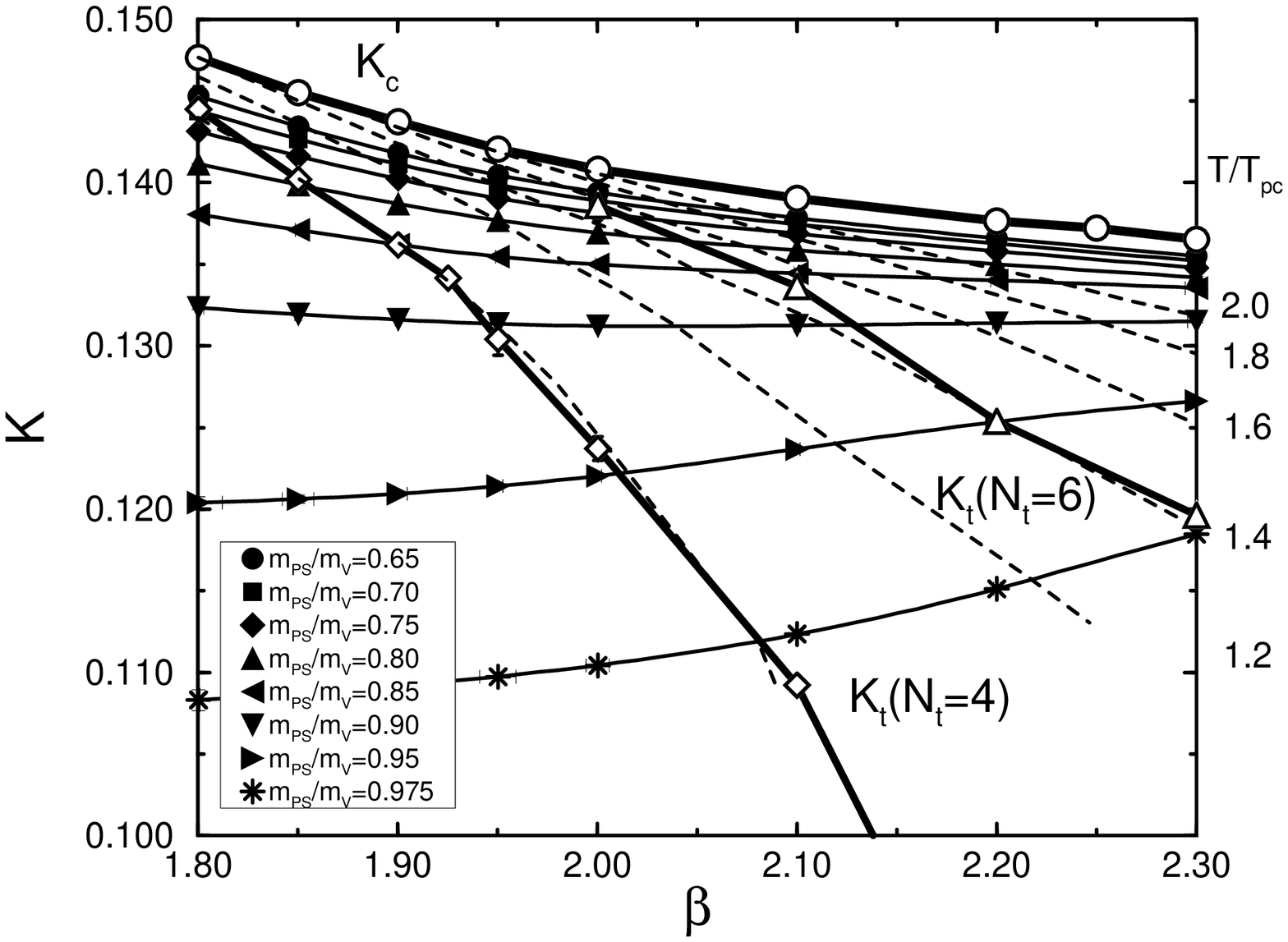}
}
\vspace{-12mm}
\caption{(a) Phase diagram and simulation points 
on $16^3 \times 4$, $16^3 \times 6$ and $16^4$ lattices.
(b) Lines of constant physics and of constant temperature.
Solid lines are $m_{\rm PS}/m_{\rm V}$ constant lines, and dashed lines 
are $T/T_{pc}$ constant lines for $N_t=4$. 
The values of $T/T_{pc}$ for the dashed lines are given 
on the right edge of the figure. 
\cite{ourPhase,ourEOS}
}
\label{fig:phase}
\end{figure}

\subsection{Lattice fermions}

Until recently, EOS with dynamical quarks has been computed only with 
the Kogut-Susskind (staggered) type lattice fermions 
because of a smaller computational demand and persistence of a part of 
the chiral symmetry on the lattice \cite{Ejiri00}.
The staggered fermion action, however, only allows four degenerate quark 
flavors, and the physically interesting cases of 2 and $2+1$ flavors are 
simulated by artificially modifying a coefficient in the update algorithm, 
effectively introducing a non-local action.
It has also been found that the critical scaling for two-flavor chiral 
transition extracted with this formalism 
does not reproduce the theoretically expected O(4) critical exponents
\cite{JLQCD98,Laermann98}. 

These problems of the staggered fermions make it imperative 
to study the issue with alternative lattice fermions --- 
Wilson-type fermions or recently proposed lattice chiral fermions. 
The latter fermions, however, require too much computational cost 
on current computers to perform a systematic investigation 
over a wide range of the parameter space. 

Wilson-type fermions have the manifest flavor symmetry and locality 
for any number of flavors.  
On the other hand, chiral symmetry is explicitly broken 
at finite lattice spacings, 
which complicates the phase diagram analysis on one hand
and introduce sizable lattice artifacts on the other hand.
Here, improvement of the lattice action was shown to be effective 
in reducing the lattice artifacts in finite temperature QCD \cite{Ejiri00}.
In particular, the expected O(4) scaling was observed 
around the two-flavor chiral transition \cite{Iwa97PRL}. 

From these observations, and also from 
a preparatory study of improved actions \cite{compara}, 
we adopt an RG improved gauge action 
coupled with a clover-improved Wilson quark action, 
as in our zero-temperature full QCD simulations 
\cite{CPPACSfull}. 

\subsection{Phase diagram and lines of constant physics}

Our simulation parameters are summarized in \cite{ourPhase,ourEOS}.
We study $N_t=4$ and 6 lattices. 
The temperature is given by $T=1/N_t a$, where we determine 
the lattice spacing $a$ by the vector meson mass $m_{\rm V}$ at $T=0$. 
Results for the phase diagram, 
together with the simulation points for the EOS study, 
are summarized in Fig.~\ref{fig:phase}(a). 
The horizontal axis is the inverse bare gauge coupling $\beta=6/g^2$. 
The vertical axis, the hopping parameter $K$, corresponds to the freedom 
of the bare quark mass.
The chiral limit is given by the $K_c(T=0)$ line.
The triangular region to the left, surrounded by the lines $K_c(N_t=4)$, 
is the parity-flavor broken phase \cite{aoki,AUU} for $N_t=4$.
Dashed lines $K_t$ are the finite temperature pseudocritical lines 
for $N_t=4$ (left) and 6 (right).
which are determined by the peak position of the susceptibility 
for a Z(3)-rotated Polyakov loop. 

In previous studies of EOS with staggered-type quark actions, 
the pressure and energy density are determined as functions of
temperature for a fixed value of bare quark mass $m_q^{bare}a$ 
or $m_q^{bare}/T = m_q^{bare}N_t a$.
While $m_q^{bare} a$ and $N_t$ are practically easy to set in simulations, 
the results at different temperatures represent values for different 
physical systems with different renormalized quark masses. 
This is not useful for phenomenological applications; 
we want to see the temperature dependence of thermodynamic observables 
for a fixed physical system with fixed renormalized quark mass, 
{\it i.e.,} on a line of constant physics. 

Therefore, it is important to evaluate the lines of constant physics
also in finite temperature physics. 
Our results for the lines of constant physics, 
which are defined by fixed values of the ratio 
$m_{\rm PS}/m_{\rm V}$ of the pseudo-scalar to vector meson masses 
at zero temperature, 
are summarized in Fig.~\ref{fig:phase}(b) by solid lines.
In the same figure, we also plot the lines of constant temperature
normalized by the pseudocritical temperature $T_{pc}$ at $K_t(N_t=4)$ 
on the same line of constant physics. 

\section{RESULTS}

\subsection{Chiral transition temperature}

We first study the chiral transition temperature $T_c$ 
in the limit of massless quarks.
In the phase diagram, Fig.~\ref{fig:phase}(a),
the chiral transition point is the crossing point of the chiral limit 
line $K_c(T=0)$ and the finite temperature pseudocritical line $K_t$.
We confirmed that the O(4) critical scaling, expected from an 
effective sigma model, is well satisfied with our data \cite{ourPhase}. 
This implies that the chiral transition is second order in two-flavor QCD.

From a theoretical argument, we expect that the parity-broken phase at 
finite $N_t$ locates in the low temperature phase \cite{AUU}. 
This provides us with an estimate of the lower bound for  
the chiral transition point; 
$\beta_{ct} \geq 1.538(46)$ from the $K_c(N_t=4)$ lines 
in Fig.~\ref{fig:phase}(a).
We can also estimate $\beta_{ct}$ using the O(4) scaling properties 
of observables:
From an O(4) fit to the chiral condensate, we find 
$\beta_{ct} = 1.469(73)$.
On the other hand, from an O(4) scaling ansatz to the $K_t$ line, 
we obtain $\beta_{ct} = 1.557(28)$.
Collecting all the results, and converting to the physical units using 
the vector meson mass at $T=0$, we obtain
$T_c = 171 \pm 4$ MeV as our best estimate \cite{ourPhase}.

\subsection{Equation of state}

\begin{figure}[tb]
\centerline{
\epsfxsize=8cm \epsfbox{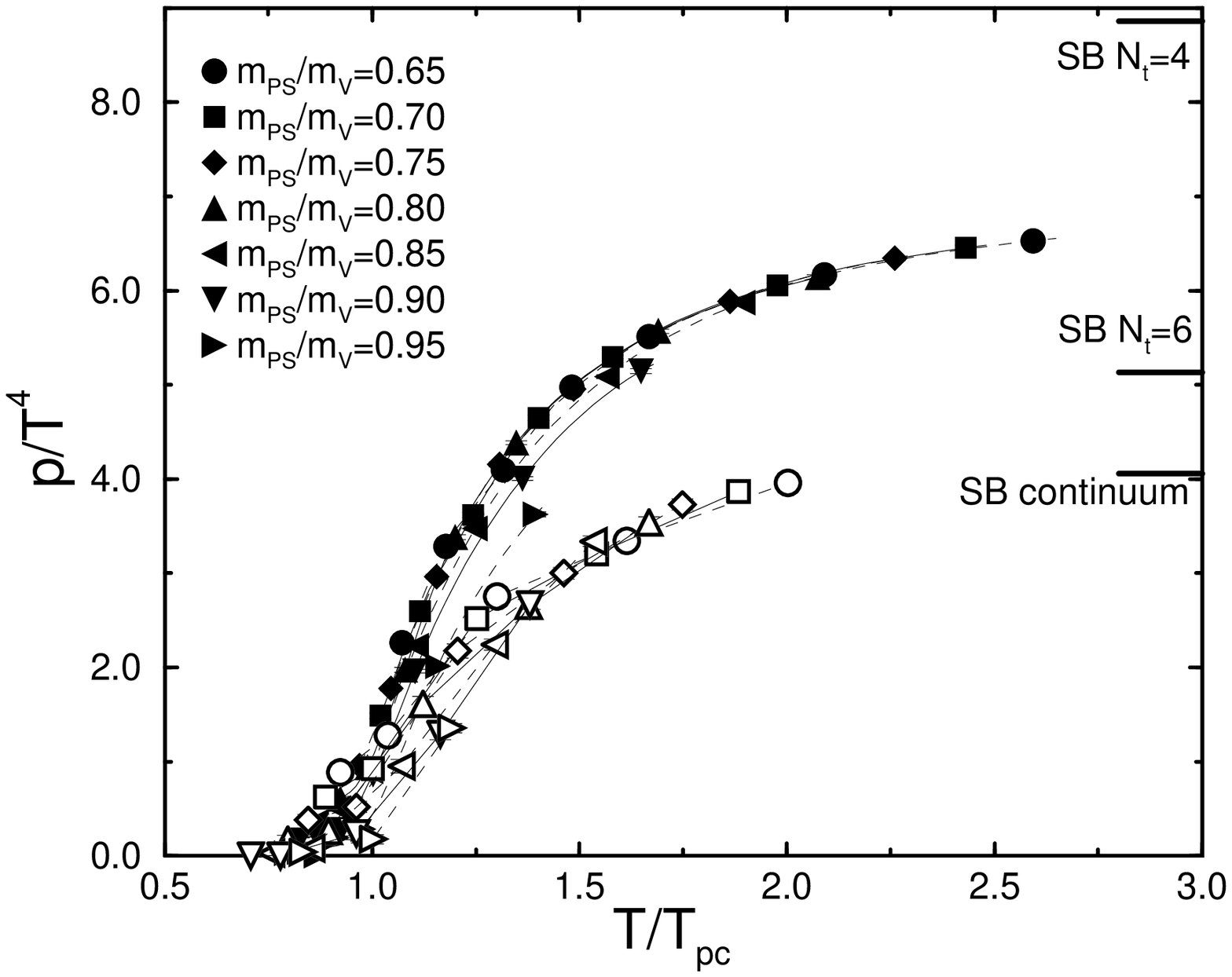}
\epsfxsize=8cm \epsfbox{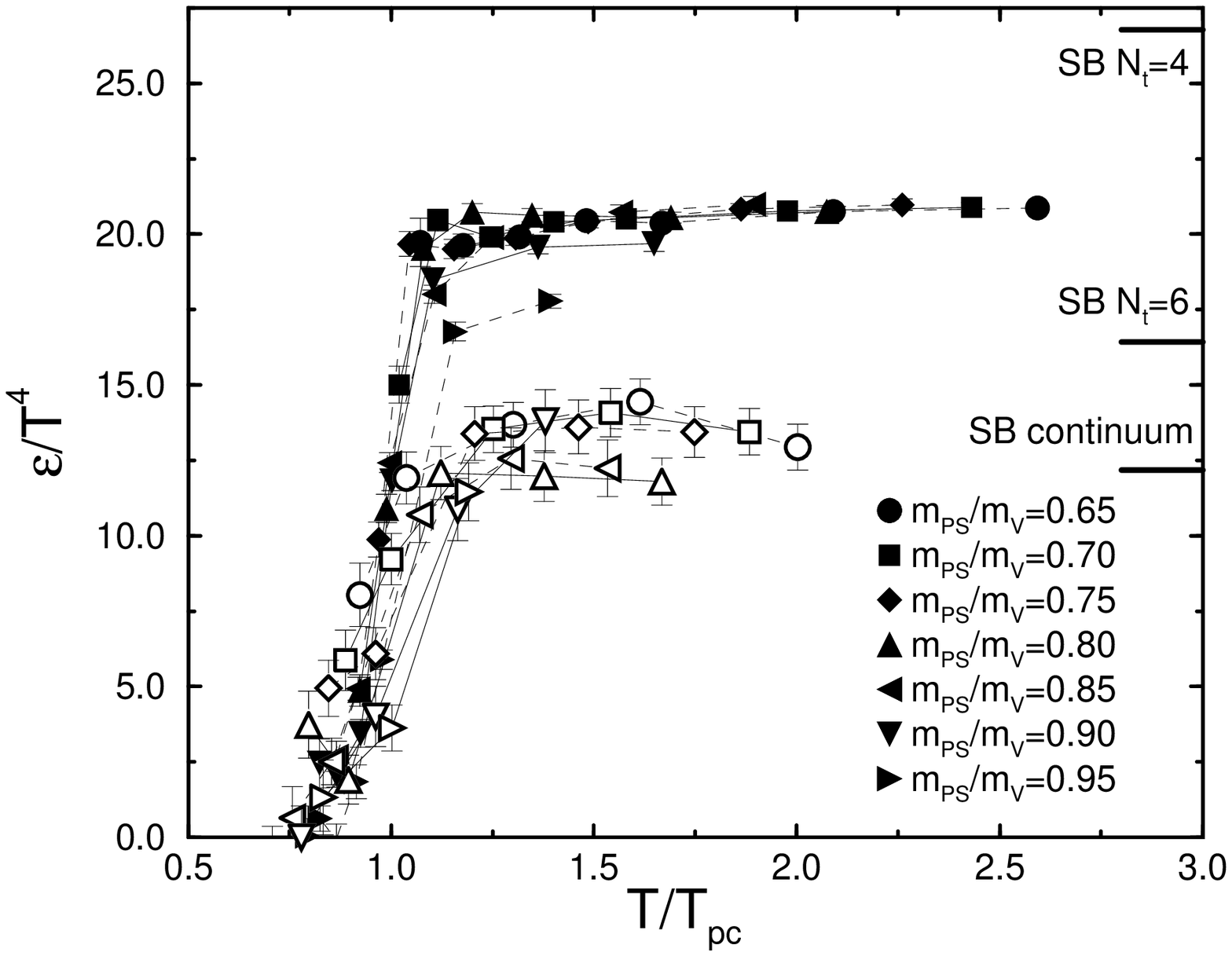}
}
\vspace{-12mm}
\caption{Pressure (a) and energy density (b) 
on $16^3\times4$ (filled symbols) and $16^3 \times 6$ (open symbols) 
lattices, as functions of $T/T_{pc}$ \cite{ourEOS}.
}
\label{fig:EOS}
\end{figure}

We calculate the pressure $p$ by the integral method \cite{Eng90}.
For the energy density $\epsilon$, we combine the results for 
$p$ and those for the interaction measure $\epsilon-3p$ 
obtained by the differential method.
%
Our results are shown in Fig.~\ref{fig:EOS}. 

The continuum limit corresponds to the limit of large $N_t$. 
Our data show a 50\% decrease from $N_t=4$ to 6, both in the 
pressure and energy density, which is too large to attempt a continuum 
extrapolation.  On $N_t=6$ lattices, however, the magnitude and temperature 
dependence of the two quantities are quite similar 
between our improved Wilson quark action and the staggered quark action
\cite{Ber97b}. 
Together with the fact that the $N_t=6$ results are close to the continuum 
SB limit at high temperatures, the approximate agreement of EOS between 
two different types of actions may be suggesting that 
the $N_t=6$ results are not far from the continuum limit. 

In Fig.~\ref{fig:EOS}, 
different values of $m_{\rm PS}/m_{\rm V}$ correspond to 
different renormalized light quark masses. 
We find that
the dependence on the quark mass is quite small for 
$m_{\rm PS}/m_{\rm V} \approx 0.65$--0.8. 
A weak quark mass dependence appears only at 
$m_{\rm PS}/m_{\rm V} \simgt 0.9$ for $N_t=4$ 
(the errors for the $N_t=6$ data are still large to conclude a quark mass 
dependence).
We find that the pressure and energy density in the quenched QCD 
($m_{\rm PS}/m_{\rm V} = 1$) are much smaller \cite{okamoto};
only about 1/7 of the values at $m_{\rm PS}/m_{\rm V} \sim 0.7$.
See Figs.~15 and 17 of \cite{ourEOS}.
These results suggest that most dynamical quark effects are saturated 
between $m_{\rm PS}/m_{\rm V} = 1$ and 0.9.
Therefore, although the light u,d quark mass point,
$m_{\rm PS}/m_{\rm V} = m_{\pi}/m_{\rho} = 1.8$,
is still far away, our values for the EOS 
at $m_{\rm PS}/m_{\rm V} \sim 0.65$--0.8 may be already 
close to those at the physical point, 
except in the very vicinity of the chiral transition point 
where a singular limit according to the O(4) critical
exponents is expected.

This result of small quark mass dependence may not be surprising 
since hadron mass results in our zero-temperature simulations 
\cite{CPPACSfull} show that the renormalized quark mass at $\mu=2$~GeV 
in the $\overline{MS}$ scheme at $m_{\rm PS}/m_{\rm V}\approx 0.65$--0.8 
equals 
$m_q^{\overline{MS}}(\mu=2 \mbox{GeV})\approx 100$--200~MeV, which is 
similar in magnitude to the critical temperature $T_c\approx 170$~MeV 
estimated for two-flavor QCD.  
For comparison, finite mass corrections 
for free fermion gas only amount to 7\% when the temperature equals 
the fermion mass $m_f$, and exceed 50\% only when $m_f/T \simgt 3$.

\section{CONCLUSIONS}

We have shown that a study of finite temperature QCD with Wilson-type quarks 
is feasible when improved lattice actions are adopted.
Based on a systematic determination of the lines of constant physics,
we calculate thermodynamic quantities in improved lattice QCD 
with two flavors of Wilson-type quarks on $N_t=4$ and 6 lattices, 
in the range $m_{\rm PS}/m_{\rm V} = 0.65$--0.95 and $T/T_c \simlt 2.5$.
We found that the quark mass dependence is small for 
$m_{\rm PS}/m_{\rm V} \simlt 0.8$. 

Concerning the continuum extrapolation, we find large scaling violations
in EOS at $N_t=4$, while $N_t=6$ results are suggested to be not far from 
the continuum limit. Therefore, a precise continuum extrapolation may be 
possible from simulations at $N_t\simgt 6$. 
This is, however, computationally quite demanding. 
In this connection, I would like to introduce our recent study of EOS 
using anisotropic lattices \cite{ourAniso}:
We found that, in quenched QCD, anisotropic lattice is efficient to reduce
discretization errors in EOS, such that a precise continuum extrapolation
is possible with much smaller amount of computer time. 
Anisotropic lattices may help performing continuum extrapolations 
also in QCD with dynamical quarks.

The studies presented in this paper are performed by the CP-PACS Collaboration.
This work is in part supported by the Grant-in-Aid
of Ministry of Education, Science and Culture
(Nos.\ 12304011 and 13640260 ).


\end{document}